*“This document is the Accepted Manuscript version of a Published Article that appeared in final form in Martín-García, B.; Velázquez, M.M. Nanoparticle Self-assembly assisted by Polymers: The Role of Shear Stress in the Nanoparticle Arrangement of Langmuir and Langmuir-Blodgett Films. Langmuir 2014, 30 (2), 509-516, Copyright © 2013 American Chemical Society. To access the final published article, see: [https://doi.org/10.1021/la404834b](https://doi.org/10.1021/la404834b).”*

**Nanoparticle Self-assembly Assisted by Polymers: The Role of Shear Stress in the Nanoparticle Arrangement of Langmuir and Langmuir-Blodgett Films.**

Beatriz Martín-García and M. Mercedes Velázquez*

*Departamento de Química Física, Facultad de Ciencias Químicas. Universidad de Salamanca, E-37008-Salamanca, Spain*

| | |
|---|---|
| Corresponding author: | M. Mercedes Velázquez |
| Corresponding address: | Departamento de Química Física<br>Facultad de Ciencias Químicas<br>Universidad de Salamanca<br>Plaza de los Caídos s/n<br>37008 Salamanca, SPAIN |
| Fax: | 00-34-923-294574 |
| E-mail: | mvsal@usal.es |

## Abstract

We propose to use the self-assembly ability of a block copolymer combined with compression-expansion cycles to obtain CdSe quantum dots (QDs) structures of different morphology. The methodology proposed consists in transferring onto mica mixed Langmuir monolayers of QDs and the polymer poly (styrene-co-maleic anhydride) partial 2 buthoxy ethyl ester cumene terminated, PS-MA-BEE, previously sheared by 50 compression-expansion cycles. Results indicate that the shear stress takes out nanoparticles at the air-water interface from metastable states and promoted a new equilibrium state of the Langmuir monolayer, then it was transferred onto mica by the Langmuir-Blodgett (LB) methodology and the morphology of the LB films was analyzed by Atomic Force Microscopy and Transmission Electron Microscopy measurements. Our results show that when the amplitude strain increases the QDs domain size decreases and the LB film becomes more ordered. The dynamic of the monolayer relaxation after cycling involves at least three timescales which are related to the damping of surface fluctuation, raft rearrangement and component movements inside each raft. Brewster Angle Microscopy allowed visualizing *in situ* the raft rearrangement at the air-water interface.

**Keywords**

Polymer, CdSe quantum dots, Langmuir-Blodgett films, Atomic Force Microscopy, Transmission Electron Microscopy, Shear stress.

## 1. Introduction

In recent years semiconductor nanocrystals have received great attention because exhibit unique size-dependent photoluminescence which are a result of quantum confinement effect and the enormous high specific surface area.[1, 2] Accordingly, they can be used in a vast spectrum of technological fields such as optics, electronics or biomedicine. Once prepared, nanoparticles have a great tendency to agglomerate[3] owing to the presence of a great deal of highly active surface atoms. The 3D-aggregation deteriorates the nanocrystals properties decreasing the quality of devices in which they are main components.[4, 5] Up to now, numerous approaches have been developed to avoid the undesirable 3D aggregation. One of the most efforts involves the use of polymer or surfactant molecules. These molecules can also be used as linkers to assemble nanoparticles into various architectures rending nanocomposites of different properties.[6] Despite the great interest generated in recent years on this topic, more work must be carried out to develop multifunctional materials with novel electric, magnetic or optical properties.[7]

For several technological applications it is necessary to support nanoparticles onto solid substrates.[8, 9] In these situations, the organization into ordered arrays leads to new interesting properties, which are strongly affected by the film morphology.[10-14] Several techniques have been used to integrate these materials onto novel devices. These techniques must achieve a great control on the deposition procedure to promote good-quality nanocomposites.[15-19] Lithographic techniques have been employed to generate well ordered nanostructures;[20-22] however, these methods are often limited because they require specialized equipment. Layer by layer deposition (LbL) is other alternative approach to self-assembly polymeric nanocomposites;[23, 24] however, it can be only used for charged polymers and hydrophilic nanoparticles, which exclude a wide number of functional materials. The Langmuir-Blodgett (LB) methodology, based on the transfer process of Langmuir films from the air-water interface onto solids, has proved to be a versatile and interesting tool to obtain water-insoluble thin films. The LB technique allows the continuous variation of particle density, spacing and arrangement by compressing or expanding the film using barriers.[25, 26] Consequently, it offers the possibility of preparing reproducible films with the control of interparticle distance necessary to exploit the nanocomposites in technological applications.[27-29] Using this methodology we have prepared LB films of self-assembled CdSe QDs and anhydride maleic polymer derivatives such as poly (octadecene-comaleic anhydride), (PMAO); or the block copolymer, poly (styrene-co-maleic anhydride) partial 2 buthoxy ethyl ester cumene terminated, PS-MA-BEE, and the Gemini surfactant ethyl-bis(dimethyl octadecylammonium bromide). Our results demonstrated that it is possible to modulate the architecture of QDs self-assemblies by modifying the composition and the surface state of the Langmuir monolayers precursors of LB films.[30-33] Moreover, film properties such as photoluminescence emission, depend on the morphology of the QD domains[32]. Typically, in these experiments, the different states of Langmuir monolayers are achieved by continuous compression of polymer/QDs or surfactant/QDs mixtures initially deposited at the air-water interface and the surface pressure value is employed to define the equilibrium properties of the different states. However, it has been reported that in several polymer[34, 35] or nanoparticle monolayers[28], the surface state achieved by compression does not correspond to the equilibrium value. This is due to different dynamic processes at the air-water interface.[25, 34-37] In these situations the metastable states often present space-filling defects that are subsequently transferred onto solids by the LB methodology decreasing the film quality. To detect this behavior it is necessary to compare the surface pressure isotherms prepared by two different ways, addition and compression methodologies. In the former, the surface pressure was continuously monitored after successive addition of different volumes of the spreading solution and the equilibrium value is taken when the surface pressure, $\pi$, remains constant at least 10 min. In the latter, the monolayers are symmetrically compressed by moving two barriers under computer control after spreading the solution. When the isotherms obtained for both methods, addition and continuous compression, do not agree with each other, the monolayer obtained for compression is far from the equilibrium state.[34, 35] This behavior is observed even for infinitely slow compression of monolayers.[38]

Several strategies such as thermal or vapor annealing have been proposed to take out nanoparticles from metastable states.[39, 40] An alternative approach consists to apply successive compression-expansion cycles to the monolayer [41] This methodology has demonstrated to provide more homogeneous and ordered monolayers[41] in which the domain morphology can be tuned by modifying the shear amplitude.[42] According to this information and taking into account that we have prepared our QD/polymer films by transferring monolayers prepared by continuous compression, we expect that some defects observed by different techniques[30,31,33] can be due to metastable states. To clarify this issue we analyze the effect of successive compression-expansion cycles on the morphology of PS-MA-BEE/QDs Langmuir and Langmuir-Blodgett films. We choose this polymer because it has been proposed among other polystyrene derivatives as a good candidate[43, 44] to fabricate sub-micrometric electronic devices due to its mechanical rigidity and good adhesion on solids[45].

We show that the shearing process renders more compact and ordered QDs domains than those monolayers prepared without shearing. Finally, the analysis of the relaxation curves obtained after stopping shear allows obtaining information of the dynamics processes at the interface. Three different relaxation processes were found and were attributed to the damping of surface fluctuation, movements of rafts constituted by QDs or QDs/polymer self-assemblies and to movements inside rafts, respectively.

## 2. Experimental section

### 2.1 Materials

Trioctylphosphine (TOP, technical grade, 90%), cadmium oxide powder (99.99%), selenium powder (99.99%), oleic acid (technical grade, 90%) and 1-octadecene were purchased from Sigma-Aldrich. The polymer poly (styrene-co-maleic anhydride) partial 2 buthoxy ethyl ester cumene terminated, PS-MA-BEE, Scheme 1, was supplied from Sigma-Aldrich. The ester:acid ratio 1:1 and the polymer molecular weight $M_n$ = 2.5 kDa were provided by the manufacturer. We use the materials as received without further purification. Chloroform (PAI, filtered) used to prepare the spreading solutions was from Sigma-Aldrich. The concentrations range for the components in the spreading solutions were from $1.25\cdot10^{-7}$ to $6.3\cdot10^{-7}$ M and 0.060 to 0.0002 mg $mL^{-1}$ for QDs and PS-MA-BEE, respectively, depending on the mixture mole ratio. The water used as subphase was ultra purified by using a combination of RiOs and Milli-Q systems from Millipore. The LB substrate mica quality V-1 was supplied by EMS (USA). The mica surface was freshly cleaved before use.

The hydrophobic CdSe QDs were synthesized by the method proposed by Yu and Peng.[46] QDs were collected as powder by size-selective precipitation with acetone and dried under vacuum. The diameter of the QDs (3.55 ± 0.05 nm) was determined by the position of the maximum of the visible spectrum of the QDs dispersed in chloroform.[47] The concentration of nanocrystals was calculated from the UV-Vis absorption spectrum of the QDs solutions by using the extinction coefficient per mole of nanocrystals at the first excitonic absorption peak.[47] UV-Vis absorption spectra were recorded on a Shimadzu UV-2401PC spectrometer.

## 2.2 Langmuir and Langmuir-Blodgett experiments

The oscillatory barrier experiments were carried out using a Langmuir Mini-trough (KSV, Finland) placed in an anti-vibration table. The Langmuir-Blodgett deposition was carried out in a Langmuir Standard-trough (KSV2000 System 2). Monolayers were transferred at constant surface pressure by symmetric barrier compression (5 mm $min^{-1}$) with the substrate into the trough by vertically dipping it up at 5 mm $min^{-1}$. Spreading solution was deposited onto the water subphase with a Hamilton microsyringe. The syringe precision was 1 μL. The surface pressure was measured with a Pt-Wilhelmy plate connected to an electrobalance. The subphase temperature was maintained at (23.0 ± 0.1)ºC by flowing thermostated water through jackets at the bottom of the trough and the water temperature was controlled by means of a thermostat/ cryostat Lauda Ecoline RE-106. The temperature near the surface was measured with a calibrated sensor from KSV.

In the oscillatory strain experiments the barriers of the Langmuir trough are driven in a harmonic oscillatory variation of the interfacial area at constant frequency, ω, and the strain, $U_0$, calculated by: $U_0 = \frac{A_0 - A(t)}{A_0}$, where $A_0$ and A(t) represent the initial and the time dependent areas, respectively. The surface pressure response, π (t), presents a sinusoidal response with the same frequency of the area perturbation.[48, 49] The strain amplitude values for oscillatory experiments range from 0.04 and 0.35 and the frequency employed varied from 0.003 to 0.012 Hz.

**2.3 Brewster Angle Microscopy (BAM)**

The Langmuir monolayers were visualized with a Brewster Angle Microscope Optrel BAM 3000 from KSV equipped with a Helium-Neon laser of 10 mW (632.8 nm) which is reflected off the air-water interface at approximately 53.15º, Brewster Angle. The microscope is also equipped with a digital camera model Kam Pro-02 (768 X 494 pixels) from EHD and an objective Mitutoyo (5x).

**2.4 Atomic Force Microscopy (AFM)**

AFM images of the LB films deposited onto freshly cleaved mica were obtained in constant repulsive force mode by AFM (Nanotec Dulcinea, Spain) with a rectangular silicon nitride cantilever (Olympus OMCL-RC800PSA) with a height of 100 µm, a Si pyramidal tip (radius < 20 nm) and a spring constant of 0.73 mN $m^{-1}$. The scanning frequencies were usually in the range of 0.5 and 1.2 Hz per line. The measurements were carried out under ambient laboratory conditions with the WSXM 5.0 program.[50]

### 2.5 Transmission Electron Microscopy (TEM)

TEM images of the LB films deposited on Formvar®-carbon coated copper grids were taken with 80 kV TEM (ZEISS EM 902, Germany). The LB deposition onto copper grids was carried out at a speed up of 1 mm $min^{-1}$.

## 3. Results and Discussion.

Prior to analyze the results, the monolayers stability during the shearing procedure was checked by recording the surface pressure in successive compression-expansion cycles followed by a waiting step. An illustrative example is shown in Figure 1 for the QDs monolayer at 2.5 mN $m^{-1}$ using the strain amplitude of $U_0 = 0.35$ and a frequency of 0.006 Hz. The compression isotherms of the polymer PS-MA-BEE and polymer/QDs monolayers were previously reported and different surface states[33] were characterized. Results inthis work also demostrated that the polymer brush conformation is reached at surface pressure values above 30 mN $m^{-1.}$ When this state was transferred onto solids, hexagonal networks with QDs adsorbed on the rims were observed. In the current work we are interested of avoiding this structure, consequently the initial state and the amplitude strain for each experiment were properly selected. Using this criterion the initial surface concentrations selected correspond to surface pressure values around 2.5 and 5 mN $m^{-1}$, respectively and the strain amplitude value, $U_0$, ranges from 0.15 to 0.25. Accordingly, during the cycling process the monolayers remained in the liquid expanded state.[33]

Results in Figure 1 show that the surface pressure decreases during the cycling period and after barrier stop, the system relaxes and reaches a constant surface pressure value. Some illustrative examples of the relaxation curves are represented in Figure 2 for the PS-MA-BEE/QDs mixed monolayer at polymer mole fraction, $X_P = 0.98$. Our results show that the amplitude of response decreases during the first compression-expansion cycles for strains values above $U_0 > 0.15$ while it remains almost constant during cycling for strains below 0.15. To illustrate this behavior for $U_0 > 0.15$, we include two representative examples as inset in Figures 2a and 2b. As can be seen Figures 2a and 2b, the amplitude of the response decreases with the strain until a given cycle and then, it remains constant.. We also observe that above 45 cycles the amplitude response remains constant for all systems. It is possible to interpret this behavior by considering that after a given cycle the monolayer reaches a stable arrangement probably due to association processes induced by shear stress. This kind of association has been reported elsewhere,[41, 51, 52] however, it is necessary to discard other processes, such as monolayer dissolution in the subphase, that could present similar behavior. To elucidate this issue we analyzed the variation of the surface pressure value obtained after the oscillatory experiments with the strain amplitude and the frequency. In the experiments 50 compression-expansion cycles were applied to the monolayer and then the barriers were stopped at the last expansion cycle. The strains values selected were above 0.15 and the frequency ranged from 0.003 to 0.030 Hz. Results show that the surface pressure value reached after the barriers stop is independent of the strain amplitude (Figure S1 of the supporting information). In addition the decrease of the surface pressure after shearing is also independent of the frequency for a given monolayer, (Figure S2 of the Supporting Information,). These results point to a new equilibrium state induced by shear stress independent of both, the strain applied for $U_0 >$ 0.15 and the frequency. Consequently, it is possible to discard that the surface pressure decrease observed after the first compression-expansion cycles could be associated to monolayer dissolution, because the mass loss by dissolution depends on both, the frequency and the strain values.

We are interested to gain insight into the nature of the processes involved after the compression-expansion cycles; therefore, we analyze the relaxation curves obtained after the cycling process. The experiments were designed in two different ways. In the first one, 50 cycles of compression-expansion were applied to monolayers and the barriers were stopped at the end of the last compression, while in the second one the barriers were stopped at the end of the last expansion. Then, the surface pressure was recorded with time until it reaches a constant value. The frequency employed in all experiments was 0.006 Hz.

As can be seen in Figure 2 the relaxation curves do not follow the exponential law and can be interpreted as a sum of three exponential functions for all systems except for polymer monolayer. In this case, the relaxation curves were interpreted by a sum of two exponential functions. The relaxation times obtained in the fit procedure at different strains are collected in Table S1 in the Supporting Information. It is interesting to note that the relaxation time values are almost independent of both, the strain applied and the kind of experiment, barriers stop at the last compression or expansion steps. The average value was calculated from the values corresponding to the different strains and the error represents the standard deviation. The average values are plotted against the polymer mole fraction, $X_P$, in Figures 3a and 3b.

The fitting results suggest three relaxation processes for QDs and polymer/QD monolayers and two relaxation mechanisms for the polymer one. The relaxation time of the fastest process is independent of both, the monolayer composition and the strain applied and the average value found was (20 ± 2) s. This process has been previously reported and was assigned to damping of surface fluctuation originated for the inertia of barrier sudden-stop.[52] The second relaxation time, $\bar{\tau}_2$, was not observed for polymer monolayers and is almost independent of the monolayer composition, the average value found was (228 ± 61) s. A similar value was observed for the relaxation process of silica nanoparticle monolayers and related to rearrangement movements of nanoparticle rafts formed by particle aggregation.[52, 53] Therefore, we assign this relaxation process to rafts movements of QDs aggregates. The existence of QDs aggregation is consistent with the decrease of the surface pressure value observed in the first compression-expansion cycles. Finally, the slowest process presents relaxation times of several thousands of seconds, $\bar{\tau}_3$, and can be related to movements inside each raft.[52] The $\bar{\tau}_3$ value slightly increases as the polymer concentration increases; this fact means that movements of components inside rafts are prevented by the polymeric matrix.

The $\bar{\tau}_3$ value for the polymer monolayers is (2334 ± 534) s. This value is similar to those ascribed to segment movements of polymer molecules and observed in a widely number of polymer monolayers.[54] According to the scenario proposed a cartoon is presented in Figure 3c to illustrate the movements in nanoparticle-containing monolayers,.

We expect that the rafts movements can be visualized by BAM and with this purpose BAM images of monolayers before and after cycling were taken. Representative images are collected in Figure 4. The images presented in Figure 4 correspond to QDs Langmuir monolayer and PS-MA-BEE/QDs mixed monolayers of polymer mole fraction values $X_P$ = 0.5 and 0.98, respectively. The BAM images taken after compression and expansion steps illustrate the raft movements. To visualize them arrows in Figure 4 point to some movements. It is also interesting to note that the most compact domains were observed for mixed films of polymer mole fraction close to 0.50; similar behavior was previously observed for PS-MA-BEE/QDs LB films built from Langmuir monolayers without shearing.[33] In this system the expansion step induces cracks that propagate perpendicularly to the stress direction, see Figure S3 of the Supporting Information. When the polymer concentration is further increased until $X_P$ = 0.98, the Langmuir monolayer presents rafts more separated than the ones corresponding to mixed monolayers of lower polymer concentration, Figures 4e and 4f.

To analyze the effect of shearing on the morphology of the LB films, we have transferred Langmuir films after 50 compression-expansion cycles at two different strain amplitude values, $U_0$=0.05 and $U_0$=0.25 and at the end of the last compression step. Some TEM images are shown in Figure 5. The images show that films of QDs directly deposited onto solids present big domains, Figures 5a and 5c, and the domain size decreases as the strain amplitude increases. Magnification of these domains, Figures 5b and 5d, indicates that they are constituted by QD clusters in which the size and shape depend on the strain amplitude applied. Thus, the cluster size decreases when the strain amplitude increases from 0.05 to 0.25. In addition the cluster shape becomes more regular when the shearing stress increases, see Fig. 5d.

TEM images of QD/polymer mixed films at polymer mole fraction values of XP= 0.5 and 0.98, are collected in Figures 5e to 5h and Figures 5k to 5n, respectively. The images show circular domains smaller than those observed in films of QDs directly deposited onto the substrate. In the case of mixed films of XP= 0.5, the TEM images collected in Figures 5f and 5h allow us to observe details of the cluster structure, thus, when the strain amplitude is 0.25 the QD domains are constituted by almost isolated nanoparticles. The two-dimensional FFT diffractogram of TEM images for mixed monolayers of XP=0.50 is shown as inset in Figures 5f and 5h. The increase of the number of bright spots observed when the strain amplitude value is 0.25 implies an increase of order induced by shearing.55 We have calculated the radial distribution functions (RDF) g(r) of the particles visualized in the TEM images for films of XP = 0.50 to infer about inter-dot order. Some differences can be observed between the RDFs of films prepared by using different strain amplitude values, see Figures 5i and 5j. Thus, the RDF corresponding to films obtained by deposition after shearing at the strain amplitude value of 0.05 shows several peaks centered at $9.2 \pm 0.5$, $14.5 \pm 0.7$, $40.5 \pm 2$, $47 \pm 2$ and $54 \pm 3$ nm. The existence of these peaks in the g(r) function indicates that QDs are not uniformly separated inside domains. This fact is in agreement with the simple visual inspection of the TEM image in the Figure 5f. In contrast, the RDF corresponding to films deposited after 50 compression-expansion cycles of strain amplitude 0.25 presents two peaks centered at $10 \pm 0.6$ and $19 \pm 0.9$ nm, indicating more regular and ordered domains than films obtained at the strain amplitude of 0.05.

Finally, when the polymer concentration is further increased until $X_P = 0.98$, the QDs coverage decreases, see Figures 5k and 5m, as expected since the QDs concentration is decreased. Moreover, the TEM images show that the QD clusters become smaller when the strain amplitude increases from 0.05 to 0.25. To analyze this effect we have carried ot the statistical analysis of the domains dimensions and results are collected in Table 1. The magnification of TEM images show disorder inside the cluster for the two films, Figures 5l and 5n.

The AFM images are collected in the Figure S4 of the Supporting Information and allow determining the profile of the different films. The AFM profile is almost independent of both, the film composition and the strain amplitude applied on the Langmuir monolayer precursor of the LB film. The value found ranges from 3 to 4 nm and agrees very well with the QD diameter determined from UV-vis spectroscopy indicating that the shearing process used in the current work does not promote the 3D aggregation of QDs.

## 4. Conclusions

The results obtained in this work demonstrate that the morphology of the QD films can be modulated by combining shear stress and the ability of the polymer PS-MA-BEE to assist the nanoparticle self-assembly. Our results show that the shearing procedure reported allows controlling the 2D aggregation of QDs by modifying the strain amplitude and the polymer concentration. Thus, we proved that by increasing the strain amplitude, the QD clustering decreases and the most ordered film can be obtained from a Langmuir film of polymer composition $X_P = 0.5$ in which we applied 50 compression-expansion cycles of 0.25 strain amplitude. From the relaxation curves obtained after the cycling procedure, we reported at least three relaxation processes at the air-water interface related to the damping of surface fluctuation, raft rearrangement and component movements inside each raft, respectively. The latter becomes slower when the polymer concentration increases. We can also visualize the raft rearrangement by means of Brewster Angle Microscopy measurements. The current results together with those previously obtained[30-33] confirm the ability of the LB methodology to modulate the self-assembly of QDs.

## 5. Acknowledgments

The authors thank financial support from ERDF and MEC (MAT 2010-19727). B.M.G. wishes to thank the European Social Fund and Consejería de Educación de la Junta de Castilla y León for her FPI grant. We also thank to Ultra-Intense Lasers Pulsed Center of Salamanca (CLPU) for the AFM measurements, especially Dr. J.A. Pérez-Hernández and to the Microscopy Electron Service (Universidad de Salamanca) for the TEM measurements.

**Supporting Information.** Monolayer stability proofs at high strains and by varying the shearing frequency; BAM images of the crack formation in monolayers during shearing; Time values of the relaxation processes with the strain applied for different monolayer compositions; and AFM images of several LB films deposited after shearing. This material is available free of charge via the Internet at http://pubs.acs.org.

**Figure Captions**

**Scheme 1.** Molecular structure of poly (styrene-*co*-maleic anhydride) partial 2-buthoxy ethyl ester cumene terminated.

**Figure 1.** Oscillatory experiment results for a QDs monolayer at the surface pressure of 2.5 mN $m^{-1}$, see text. The experiment was carried out using the strain amplitude value of 0.35 and at the frequency of 0.006 Hz. The inset shows a zoom of the waiting stages pointed out with the box.

**Figure 2.** Relaxation curves of the PS-MA-BEE/QDs mixed monolayer of polymer mole fraction $X_P$ = 0.98. Lines correspond to curves calculated from multiexponential functions and parameters in Table S1, see text. The relaxation curves were obtained after 50 compression-expansion cycles and the end of the last compression (a) and of the last expansion (b). The amplitude strains were: 0.06 (a) and 0.18 (b) and the oscillation frequency was 0.006 Hz. The initial surface pressure of these experiments was 5 mN $m^{-1}$.

**Figure 3.** (a, b) Polymer mole fraction dependence of the relaxation time. See text for the experimental conditions. The relaxation time for the PS-MA-BEE monolayer is represented as star in (b) and open circles correspond to relaxation curves obtained by stopping the barriers at the end of the last expansion step. (c) The cartoon represents the movements that dominate the relaxation of the QDs monolayers.

**Figure 4.** BAM images (800x600 μm) showing several rafts movements promoted by expansion (b, d, f) after monolayer compression (a, c, e) for PS-MA-BEE/QDs

monolayers of polymer mole fraction: $X_P$ = 0 (a, b); $X_P$ = 0.50 (c, d) and $X_P$ = 0.98 (e, f). The arrows point to the visible raft movements.

**Figure 5.** TEM images of PS-MA-BEE/QDs LB films deposited after 50 compression-expansion cycles and amplitude strains of 0.05 (a, b, e, f, k, l) (left side) and 0.25 (c, d, g, h, m, n) (right side), for monolayers of different composition: $X_P$ = 0 (a-d); $X_P$ =0.50 (e-h) and $X_P$ =0.98 (k-n). The yellow boxes indicate the areas magnified. The insets in 5f and 5h show the 2D FFT diffractograms of these TEM images, see text. calculated with the WSxM 5.0 software®. RDFs calculated from TEM images of PS-MA-BEE/QDs LB films ($X_P$ =0.50) for amplitude strains of 0.05 (i) and 0.25 (j), see text for details. The initial surface pressure of the monolayers was 2.5 mN $m^{-1}$ in all cases.

## Table of Contents

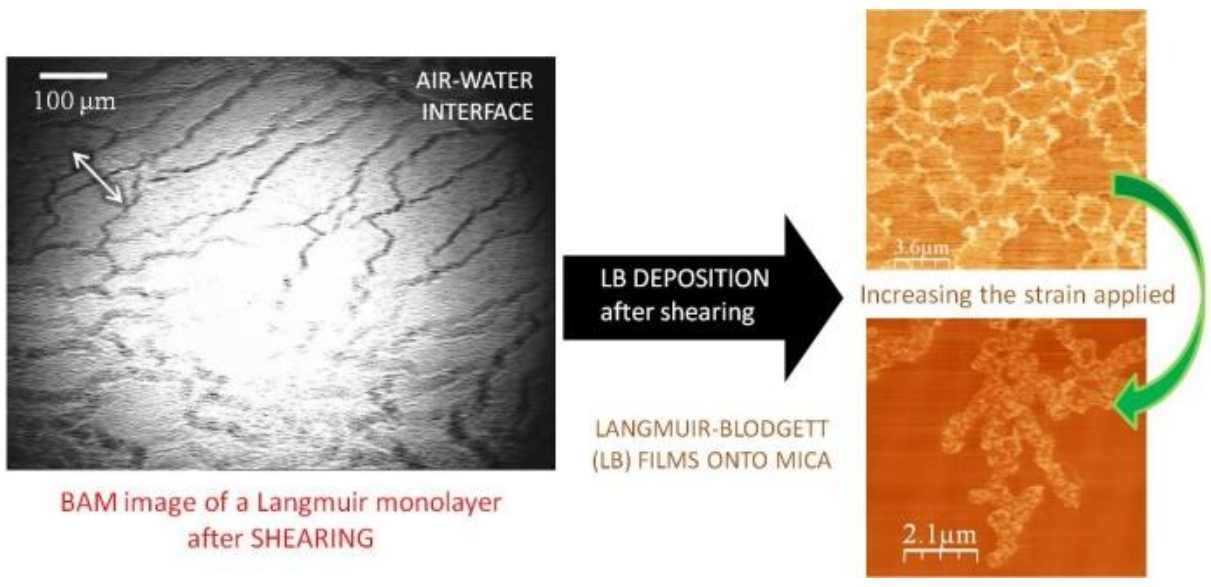

**Scheme 1.**

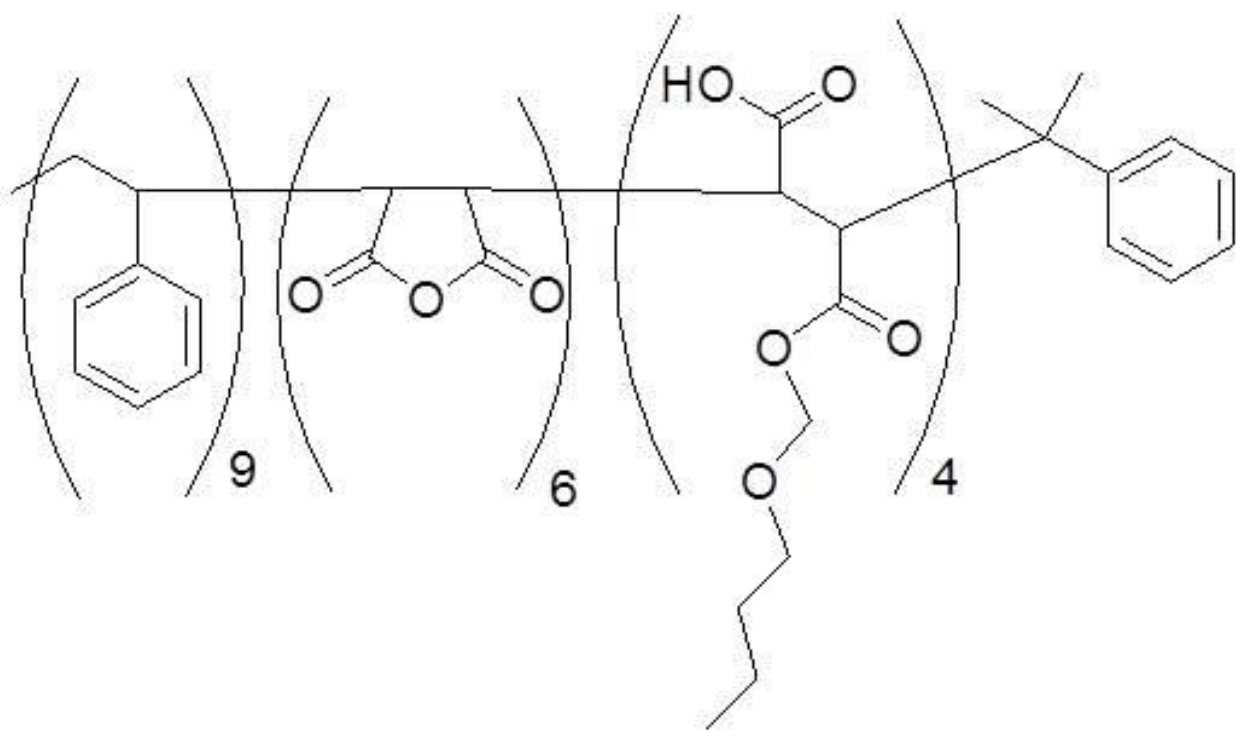

**Figure 1.**

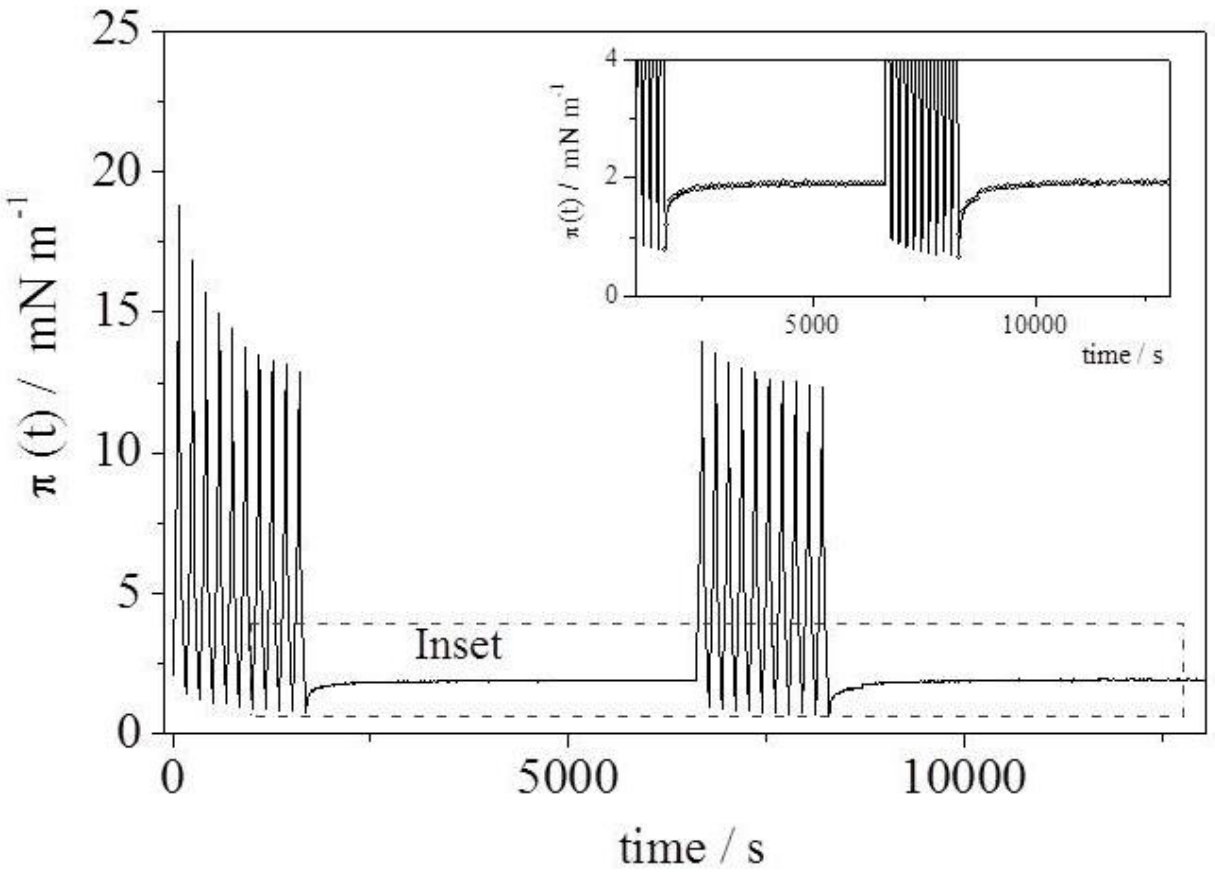

25
20
15
10
5
0
π (t) / mN m-1
Inset
0
5000
10000
time / s
4
2
0
π (t) / mN m-1
5000
10000
time / s

**Figure 2.**

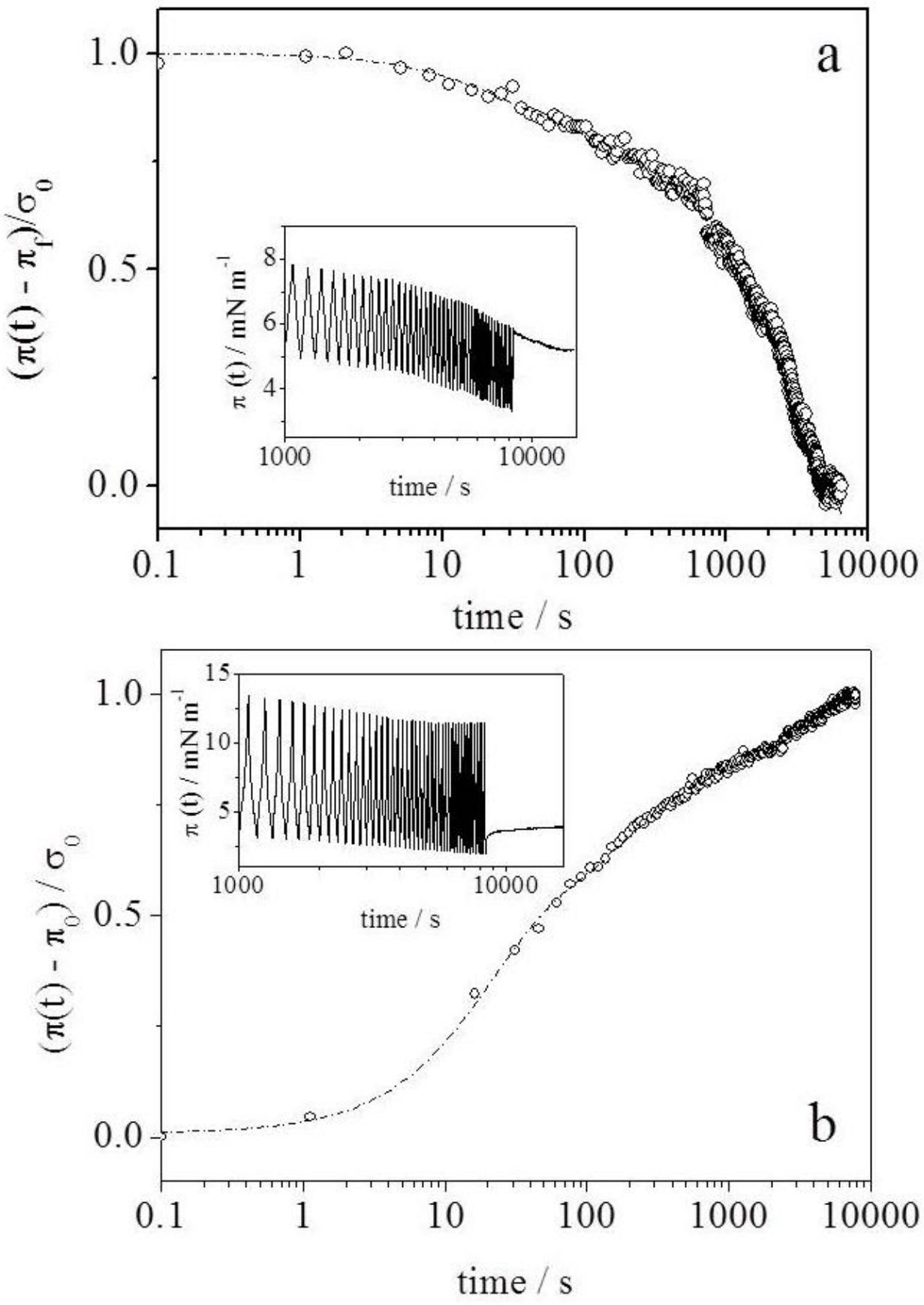
a
1.0
0.5
0.0
0.1
1
10
100
1000
10000
time / s
$(\pi(t) - \pi_f)/\sigma_0$
$\pi$ (t) / mN m$^{-1}$
8
6
4
1000
10000
time / s
b
1.0
0.5
0.0
0.1
1
10
100
1000
10000
time / s
$(\pi(t) - \pi_0) / \sigma_0$
$\pi$ (t) / mN m$^{-1}$
15
10
5
1000
10000
time / s

**Figure 3.**

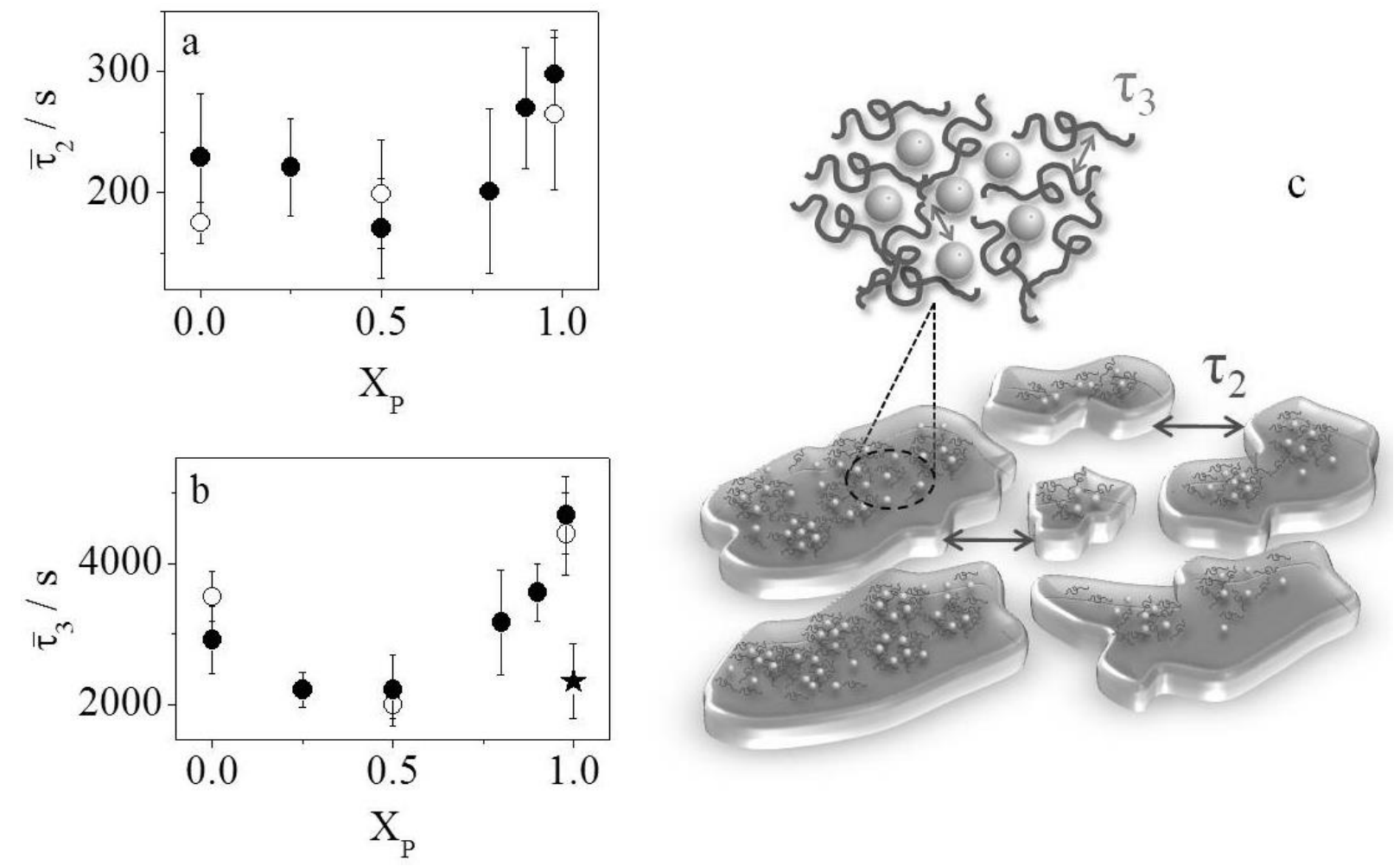
a
$\bar{\tau}_2$ / s
300
200
0.0
0.5
1.0
$X_P$
b
$\bar{\tau}_3$ / s
4000
2000
0.0
0.5
1.0
$X_P$
$\tau_3$
c
$\tau_2$

**Figure 4.**

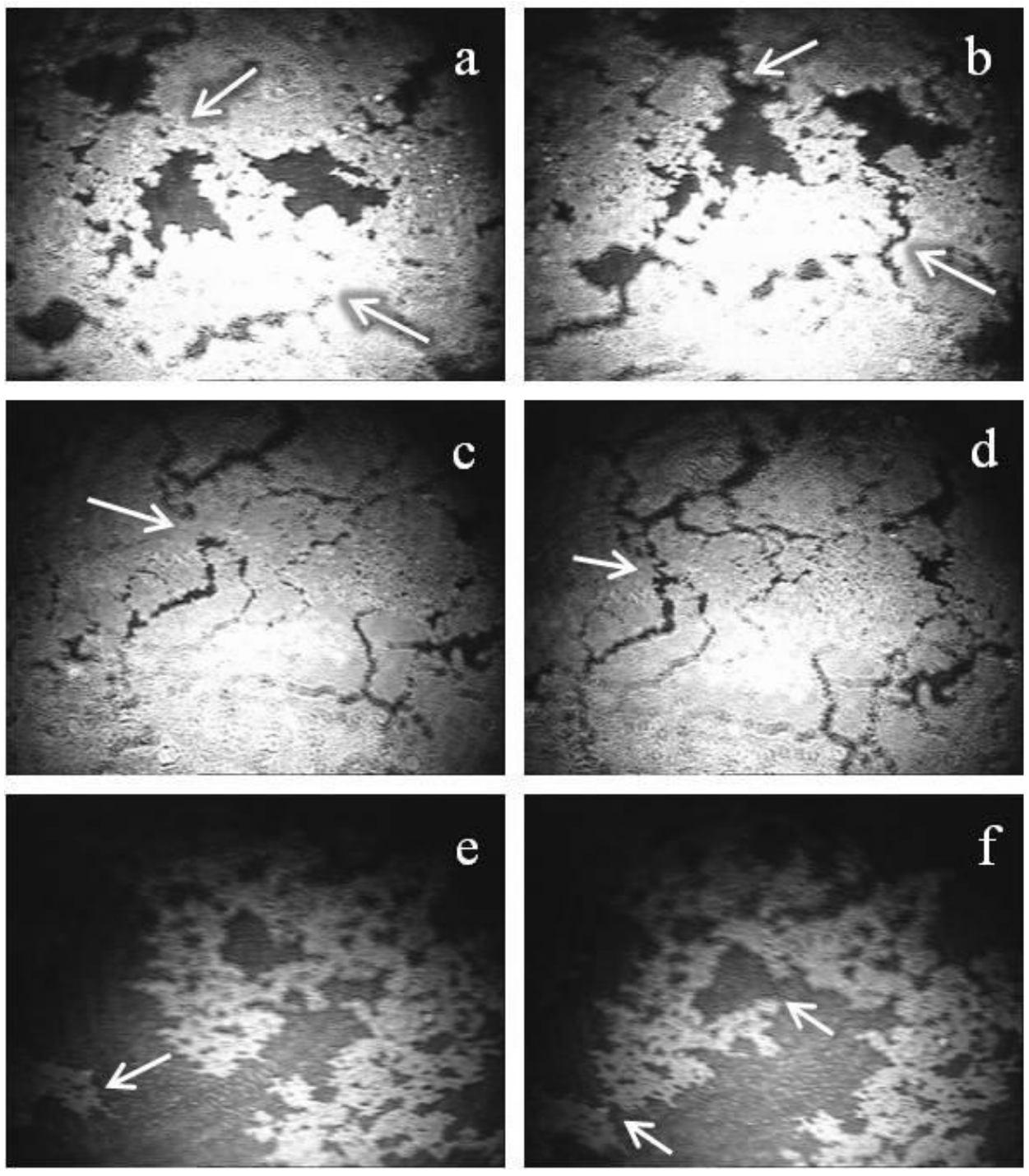

**Figure 5.**

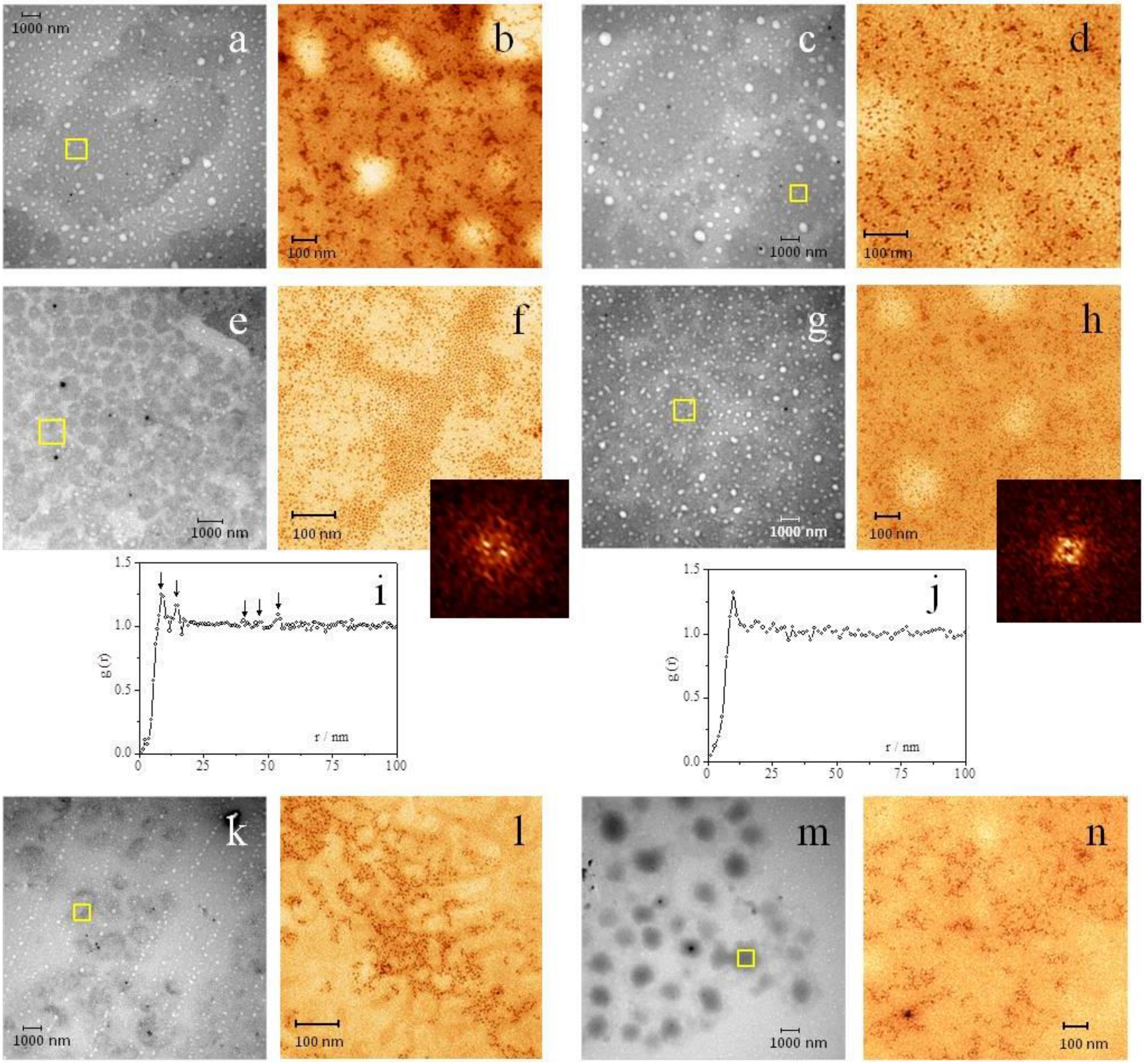

*"This document is the Accepted Manuscript version of a Published Article that appeared in final form in Martín-García, B.; Velázquez, M.M. Nanoparticle Self-assembly assisted by Polymers: The Role of Shear Stress in the Nanoparticle Arrangement of Langmuir and Langmuir-Blodgett Films. Langmuir 2014, 30 (2), 509-516, Copyright © 2013 American Chemical Society. To access the final published article, see: https://doi.org/10.1021/la404834b."*

**Table 1.** Average values of the feature dimensions of the PS-MA-BEE/QD mixed LB film prepared at $X_P$ = 0.98 obtained from AFM measurements.

| Strain ($U_0$) | X-direction / µm | Y-direction / µm |
|---|---|---|
| **0.05** | 2.53 ± 0.56 | 2.17 ± 0.27 |
| **0.25** | 0.45 ± 0.12 | 0.38 ± 0.13 |

*Error represents the standard deviation determined from at least 20 surface features.